\documentclass[10pt,twocolumn,letterpaper]{article}

\usepackage{iccv}
\usepackage{times}
\usepackage{epsfig}
\usepackage{graphicx}
\usepackage{amsmath}
\usepackage{amssymb}

\usepackage{multirow}

\usepackage[pagebackref=true,breaklinks=true,letterpaper=true,colorlinks,bookmarks=false]{hyperref}

\iccvfinalcopy 

\def\iccvPaperID{0022} 

\ificcvfinal\fi

\newcommand{\clahe}{\mbox{\textsc{CLAHE}}\xspace}
\newcommand{\dl}{\mbox{\textsc{DL}}\xspace}

\newcommand{\deepcontrast}{\mbox{\textsc{DeepContrast}}\xspace}

\def\eg{\emph{e.g}\onedot} 
\def\ie{\emph{i.e}\onedot}

\begin{document}

\title{DeepContrast: Deep Tissue Contrast Enhancement using Synthetic Data Degradations and OOD Model Predictions}

\author{Nuno Pimpão Martins$^1$\\
{\tt\small pimpaoma@mpi-cbg.de}
\and
Yannis Kalaidzidis$^1$\\
{\tt\small kalaidzi@mpi-cbg.de}
\and
Marino Zerial$^1$\\
{\tt\small zerial@mpi-cbg.de}
\and
Florian Jug$^{1,2}$\\
{\tt\small florian.jug@fht.org}
\and
\small $^1$ MPI-CBG, Dresden, Germany
\and
\small $^2$ Fundazione Human Technopole, Milano, Italy
}

\maketitle
\ificcvfinal\thispagestyle{empty}\fi

\begin{abstract}
Microscopy images are crucial for life science research, allowing detailed inspection and characterization of cellular and tissue-level structures and functions.
However, microscopy data are unavoidably affected by image degradations, such as noise, blur, or others. 
Many such degradations also contribute to a loss of image contrast, which becomes especially pronounced in deeper regions of thick samples.
Today, best performing methods to increase the quality of images are based on Deep Learning approaches, which typically require ground truth (GT) data during training.
Our inability to counteract blurring and contrast loss when imaging deep into samples prevents the acquisition of such clean GT data.
The fact that the forward process of blurring and contrast loss deep into tissue can be modeled, allowed us to propose a new method that can circumvent the problem of unobtainable GT data.
To this end, we first synthetically degraded the quality of microscopy images even further by using an approximate forward model for deep tissue image degradations. 
Then we trained a neural network that learned the inverse of this degradation function from our generated pairs of raw and degraded images.
We demonstrated that networks trained in this way can be used out-of-distribution (OOD) to improve the quality of less severely degraded images, \eg the raw data imaged in a microscope.
Since the absolute level of degradation in such microscopy images can be stronger than the additional degradation introduced by our forward model, we also explored the effect of iterative predictions.
Here, we observed that in each iteration the measured image contrast kept improving while detailed structures in the images got increasingly removed. 
Therefore, dependent on the desired downstream analysis, a balance between contrast improvement and retention of image details has to be found.
\end{abstract}

\section{Biological Motivation}

\begin{figure}[t]
\begin{center}
\includegraphics[width=0.9\linewidth]{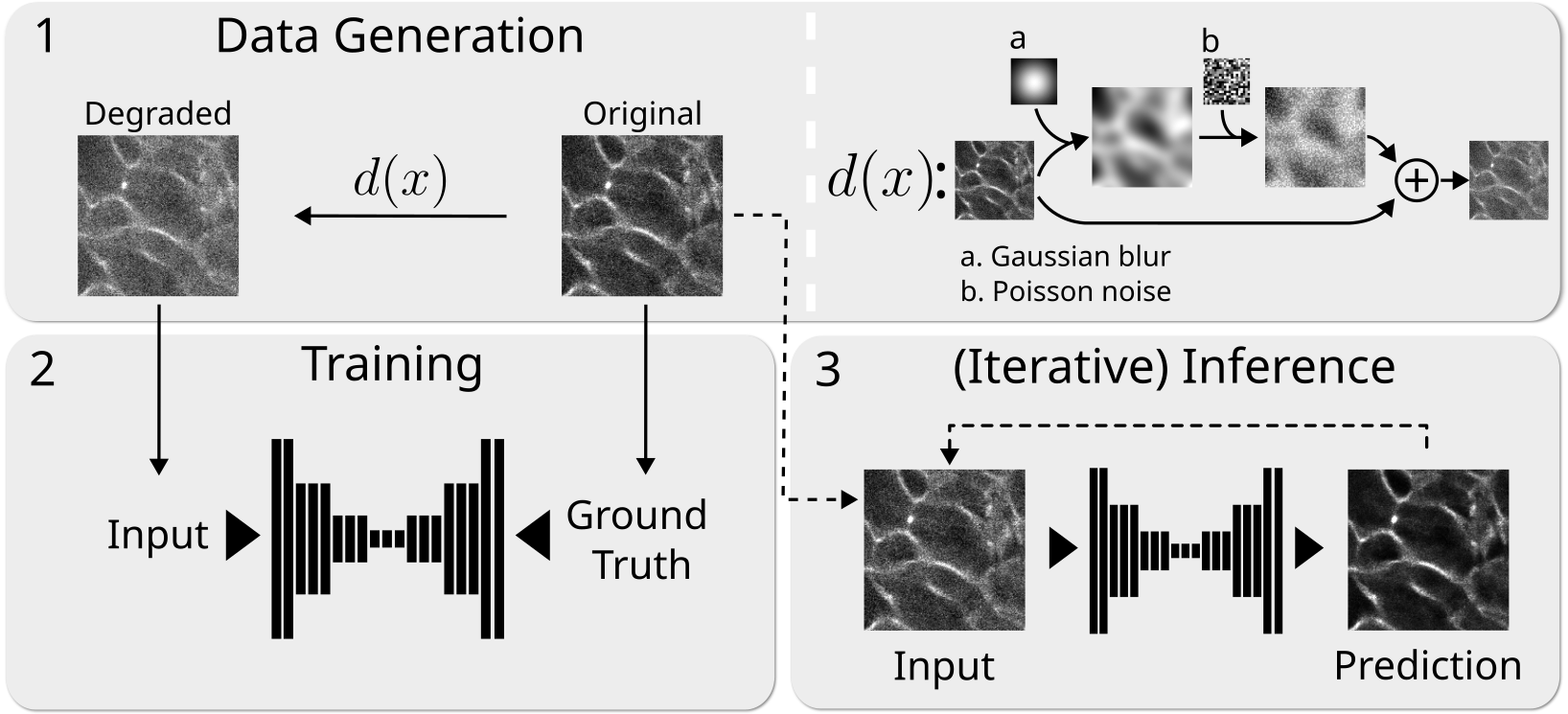}
\end{center}
    \caption{Proposed scheme to improve deep tissue contrast.
    \textbf{(1)}~Pairs of data for supervised training are generated by degrading raw microscopy images using a suitable degradation function $d(x)$ composed of a blurring and a noising step. 
    \textbf{(2)}~During supervised network training, synthetically degraded images are used as inputs and the original images as targets. 
    \textbf{(3)}~During inference, we feed the original raw microscopy images once or iteratively into the trained network (see Section~\ref{sec:methods_iterative_pred}).
    }
    \label{fig:training_scheme}
\end{figure}

In this work we applied our method (\deepcontrast) to microscopy images of liver tissue.
The liver is a frequently studied system in biomedical research, due to its vital functions in the human body, \eg blood detoxification and bile production.
Liver tissue is dense and compact, composed of many different cell types that display an intricate three-dimensional architecture.
Still, many aspects of this structure are not fully understood, which drives biomedical research to use modern microscopy techniques to image large 3D sections of liver tissue at the highest achievable quality and resolution.

Data presented in this work was obtained using a Laser Scanning Confocal Microscope (LSM).
This modality allowed us to obtain highly detailed image data in large and thick samples with sub-cellular resolution in all three spatial dimensions. 
Unfortunately, the image quality inevitably degrades in deeper layers of the imaged liver tissue, mostly due to light scattering.
This poses a challenge to better our understanding of mesoscale structures that shape the liver in its full 3D complexity.
Therefore, methods that facilitate the downstream analysis of large 3D image data are much sought after.

\begin{figure*}
\begin{center}
     \includegraphics[width=1\linewidth]{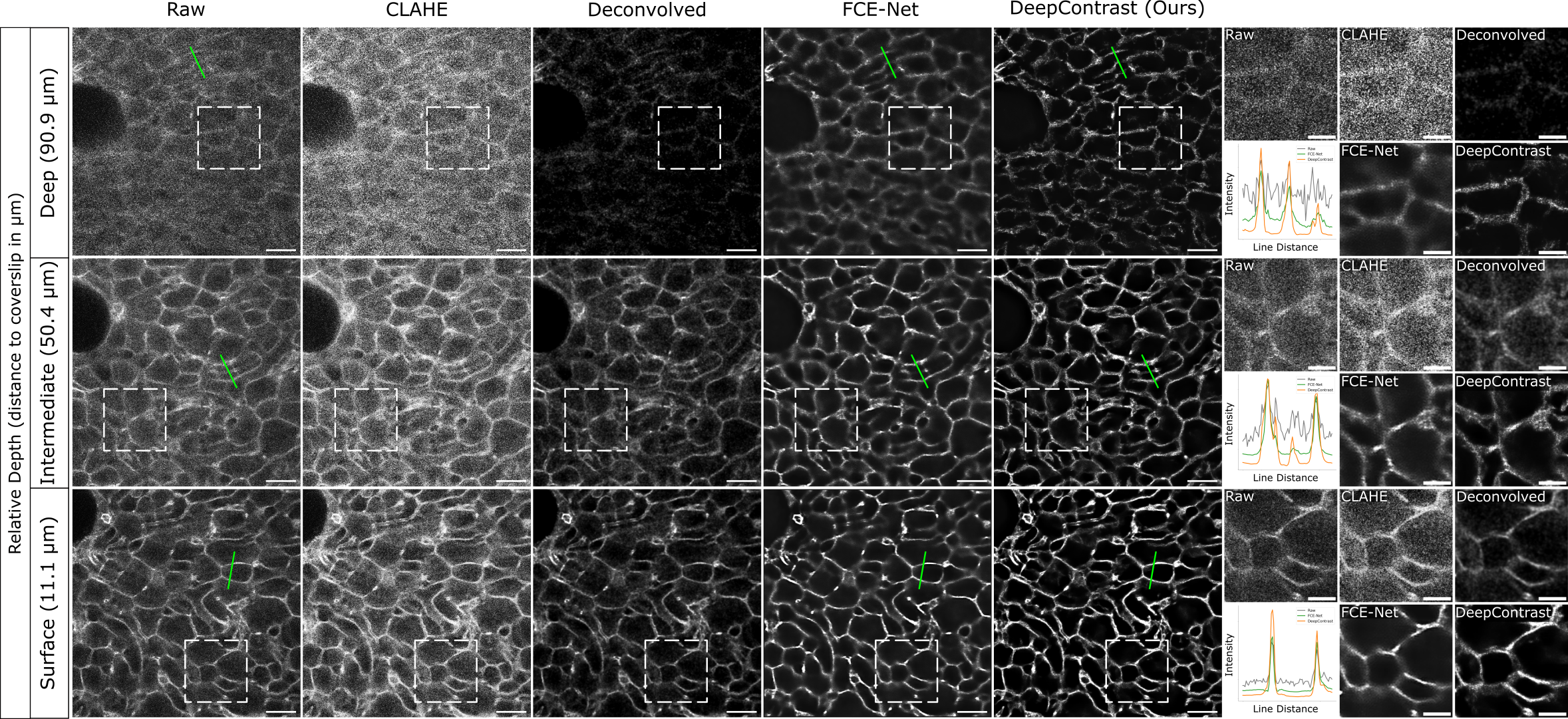}
\end{center}
    \caption{Qualitative results.
    Images of liver tissue sections stained with Phalloidin as a proxy for cell borders, used to compare our results (\deepcontrast Model~A) to several baseline methods. 
    Rows show image planes at different depths in the liver tissue.
    Columns show the raw input, results obtained with CLAHE~\cite{zuiderveld_contrast_1994}, Huygens deconvolution (see Section~\ref{sec:experiments_baselines}), best FCE-Net~\cite{zhang_fce-net_2022} results ($3\times$), and our best \deepcontrast results ($3\times$), respectively. 
    The three rightmost columns shows the inset areas marked by dashed boxes and line plots of raw intensities, the FCE-Net, and \deepcontrast (along the green line in the respective images). 
    \textit{Scale bars}: $20 \mu m$ in full size images, $10 \mu m$ in insets. 
    }
    \label{fig:results_qualitative}
\end{figure*}

\begin{figure*}
\begin{center}
    \includegraphics[width=1\linewidth]{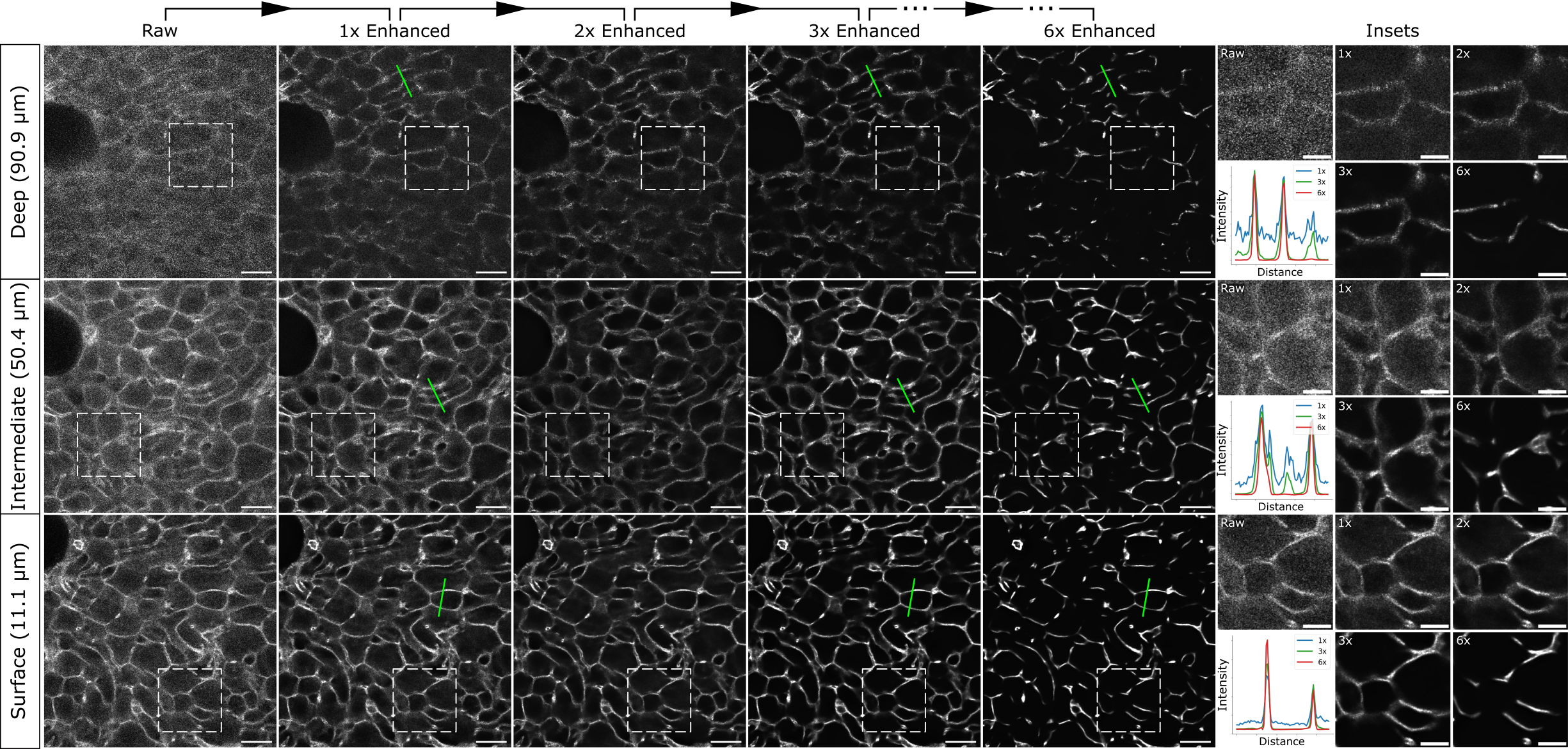}
\end{center}
    \caption{Qualitative results of iterative OOD model application. 
    Contrast of the image data of Figure~\ref{fig:results_qualitative} iteratively enhanced using a trained \deepcontrast network (Model~A).
    Rows show, analogous to Figure~\ref{fig:results_qualitative}, image planes at different depth into the imaged tissue.
    Columns show the raw input data, and the results of applying \deepcontrast a single time, two, three (same as in Figure~\ref{fig:results_qualitative}), and six consecutive times.
    The three rightmost columns shows the inset areas marked by dashed boxes and line plots along the green lines in the raw data, and along the $1\times$, $3\times$, and $6\times$ enhanced outputs.
    Note that, while contrast is continuously enhanced, too many iterative applications cause a notable loss of image details.
    \textit{Scale bars}: 20 $\mu m$ in full size images, 10 $\mu m$ in insets.
    }
    \label{fig:iterative_inference}
\end{figure*}

\begin{figure}[ht]
\begin{center}
     \includegraphics[width=1\linewidth]{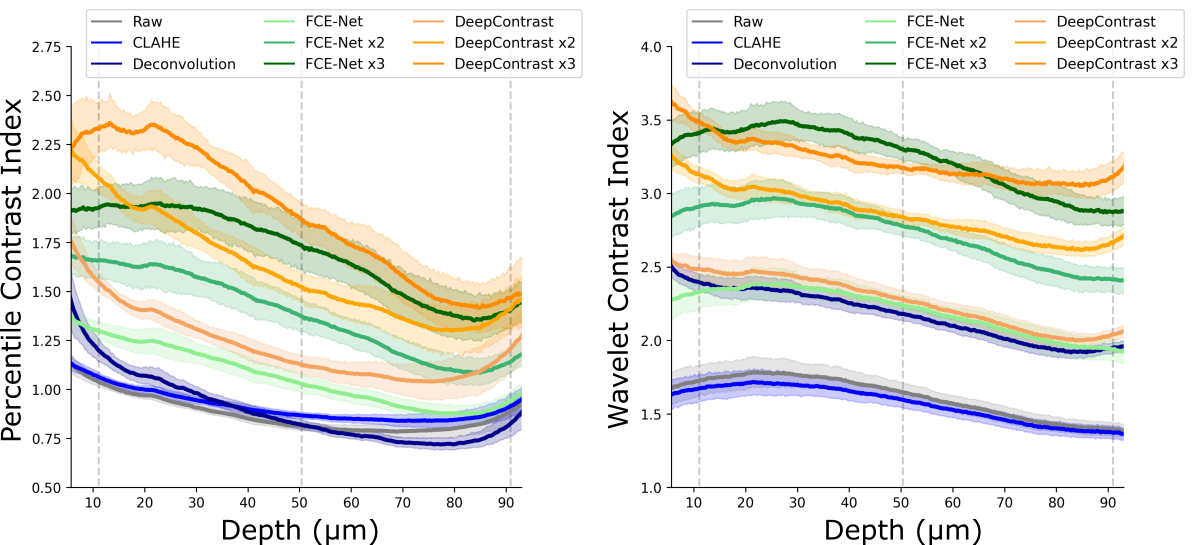}
\end{center}
    \caption{Quantitative results.
    Contrast quantification using the Percentile Contrast Index (see Section~\ref{sec:contrast_quantification}) and the Wavelet Contrast Index~\cite{albright_paintings_2023} (higher values are better) represented as average and $95\%$ Confidence Intervals at each depth ($N=18$).
    Dashed vertical grey lines depict depths shown in Figure~\ref{fig:results_qualitative}.
	Multiple iterations of FCE-Net~\cite{zhang_fce-net_2022} and our \deepcontrast (Model~A) approach show image contrast is further improved when these networks are iteratively applied.
    }
    \label{fig:results_quantitative}
\end{figure}

\begin{figure*}
\begin{center}
    \includegraphics[width=1\linewidth]{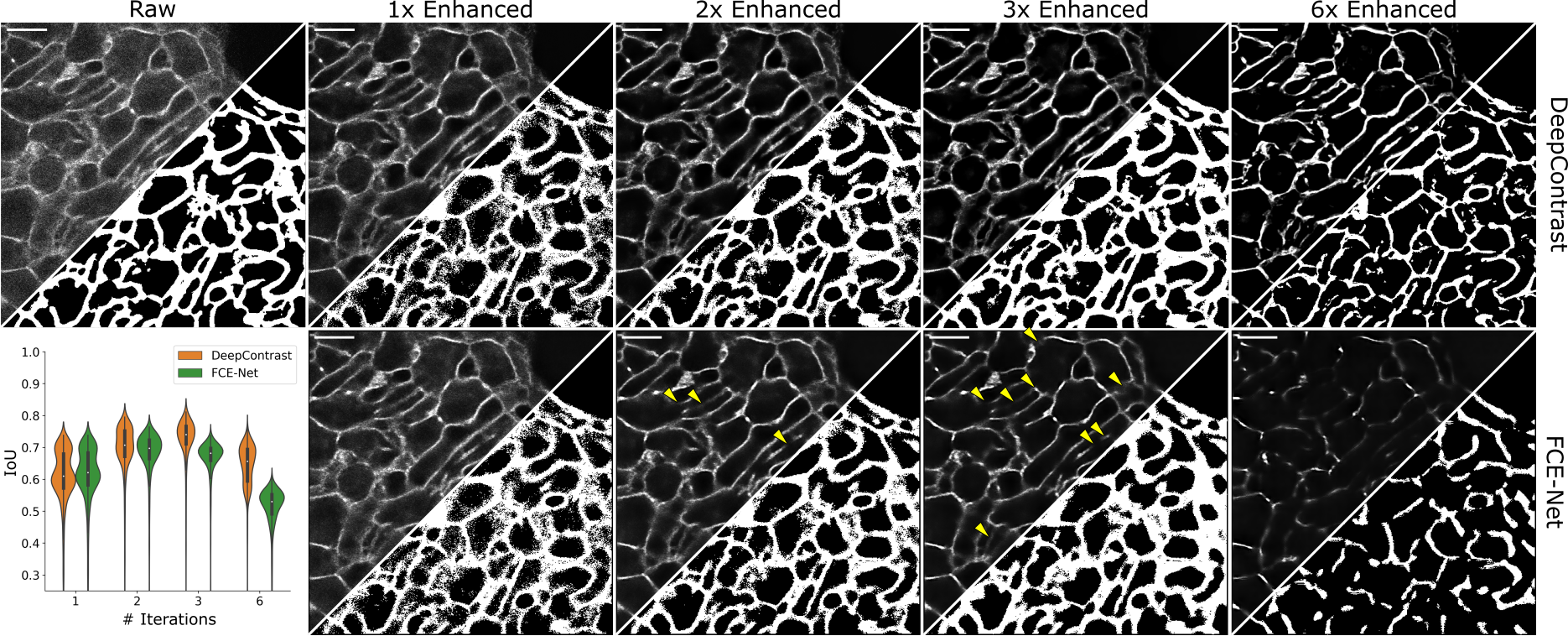}
\end{center}
    \caption{Qualitative and quantitative results of segmentation masks created at multiple iterations of contrast enhancement. 
    Left side column shows Raw input and segmentation mask (Section~\ref{sec:downstream_analysis}).
    Each further column shows different iterations of contrast enhancement and corresponding segmentation of cell borders.
    Top row shows inference results with \deepcontrast and bottom row shows inference results with FCE-Net.
    Yellow arrow heads in images highlight lost or degraded structures when comparing \deepcontrast and FCE-Net.
    Violin plots shows distribution of IoU values between the different contrast enhancement methods and raw segmentation masks used as reference at multiple iterations ($N=2682$), showing faster decrease of IoU values and more abundant mistakes in segmentation with FCE-Net.
    }
    \label{fig:multi_segmentation}
\end{figure*}

\section{Related Work}

Classical algorithms to enhance contrast in images often rely on the intensity histogram, typically altering the overall histogram landscape with a set of predefined rules to either obtain a more uniformly sampled distribution or to match an histogram obtained from a desired reference.
Examples to such algorithms are known as histogram matching~\cite{gonzalez2006digital} or histogram equalization~\cite{pizer_adaptive_1987}.
These approaches are not content-aware, \ie, they follow specific rules independent of the structures visible in the image to be modified. Contrast Limited Adaptive Histogram Equalization (\clahe)~\cite{zuiderveld_contrast_1994} is an example of widely used histogram equalization method.
We use this method as one of our baselines methods due to its popularity and widespread use in scientific image processing protocols.

Another popular family of algorithms used to improve image quality and contrast are feed-forward deconvolution methods such as the one by Richardson-Lucy~\cite{richardson_bayesian-based_1972, lucy_iterative_1974}, or the popular Huygens software from \href{https://svi.nl/}{Scientific Volume Imaging}. 
These are iterative approaches that attempt to undo the blurring induced by the point spread function (PSF) of the microscope. 
The main drawback of such approaches is the assumption of spatial invariance of the PSF, which does not hold in thick dense tissue microscopy.

Deep Learning (\dl) based applications have proven to perform especially well on several image restoration tasks like denoising~\cite{weigert_content-aware_2018, krull_noise2void-learning_2019, batson_noise2self_2019, goncharova_improving_2020, prakash_fully_2021, prakash_interpretable_2022}, deconvolution~\cite{chen_three-dimensional_2021, guo_rapid_2020, li_incorporating_2022}, and super-resolution~\cite{Ouyang2018-zc,Nehme2018-rm,tai_image_2017, zhang_image_2018, zhang_residual_2018}.

Content Aware Image Restoration (CARE)~\cite{weigert_content-aware_2018}, uses supervised \dl methods to restores microscopy image quality in various ways.
However, in order to use CARE, it is necessary to obtain low and high quality versions of the same objects and structures, which is not possible in many real-world scenarios, such as the one presented in this work.
One interesting insight with respect to out-of-distribution (OOD) denoising is presented in~\cite{mohan_robust_2020}.
In it, the authors show that a network without trainable bias terms is more robust when applied to inputs that contain levels of noise that are inconsistent (OOD) with respect to the training data.

One popular way to solve the problem of GT data being required is to synthetically generate the required training pairs~\cite{weigert_isotropic_2017}.
In~\cite{fang_deep_2021}, the authors used ``crapified'' images, as they call it, to obtain said training pairs for training super-resolution networks and networks that increase the temporal consistency in time-lapse movies.
Others have used synthetic data generation for object detection~\cite{yao_deep-learning_2020} or segmentation~\cite{dunn_deepsynth_2019}.

Work specifically concerned with enhancing image contrast is less common. 
The FCE-Net~\cite{zhang_fce-net_2022} proposes a network architecture specifically designed to enhance image contrast in biological image data.
Since this makes FCE-Net our closest competitor, despite technically being a quite different approach, we will always also compare our own results to the ones obtainable with the FCE-Net.

\section{Methods}
\label{sec:methods}
\label{sec:methods_training_scheme}
Inspired by~\cite{weigert_content-aware_2018} and~\cite{fang_deep_2021}, we also set out to use the machinery of supervised learning in deep neural networks.
In our case, for the sake of improving image contrast in microscopy data of large tissue samples.
To this end, we synthetically generate appropriate training data, \ie pairs of images that are of lower and higher contrast.
Naturally, we cannot synthetically remove scattered light and noise from raw microscopy data, otherwise the very task we are seeking a solution for would be solved already.
Instead, we can add additional light scattering and noise to the available raw data, making it even worse (see Figure~\ref{fig:training_scheme}).

More specifically, our degradation function is
\begin{equation} \label{eq:degradation}
	\text{d}(x) = \alpha \cdot x + (1-\alpha)\cdot \text{n}(\text{b}(x)),
\end{equation}
where $x$ is a raw input image, $\alpha$ a hyperparameter that controls the blending between $x$ and $\text{n}(\text{b}(x))$, $\text{b}$ is a blurring function that models light scattering in biological tissue, and $\text{n}$ a function adding noise.
In line with the most dominant noise in low-light fluorescent microscopy, $\text{n}$ is adding Poisson noise to the blurred data.

Light scattering depends on refractive index transitions throughout the sample and a precise forward model is not easy to compute.
For our purposes, the simple approximation introduced in Equation~\ref{eq:degradation} leads to contrast enhancement results that outperform existing methods like the FCE-Net (see Section~\ref{sec:experiments}).

Once a body of input data $X=(x_1,x_2,\ldots,x_k)$ is further degraded to $D=(d(x_1),\ldots, d(x_k))$, we use image pairs $(d_i,x_i)$ for supervised training of a contrast enhancement network (Model~A). 
The network was trained as a bias-free~\cite{mohan_robust_2020} U-Net~\cite{ronneberger_u-net_2015}.
The reason for \textit{not} training the bias terms of the network nodes is that, once trained, it is intended to be applied to images $x_i$ or similar, which are less severely affected by degradations, therefore, OOD with regards to $d_i$.

\subsection*{Iterative Predictions}
\label{sec:methods_iterative_pred}
Since the absolute level of degradation in such microscopy images can be stronger than the additional degradation introduced by our forward model, we also explored the effect of iterative prediction
Iterative predictions with a trained \deepcontrast network ($\text{DC}$) are simply multiple applications of $\text{DC}$ to the input $x$.
For example, the final \deepcontrast ($3\times$) prediction $y$ is computed by
\begin{equation} \label{eq:iterations}
    y = \text{DC}(\text{DC}(\text{DC}(x))).
\end{equation}

The experiments we describe below and the results we show in Figure~\ref{fig:iterative_inference} indicate that iterative predictions indeed keep increasing image contrast.
It must be noted, that better contrast does not necessarily mean that the predicted image is better for downstream analysis. 
Image faint details might get lost at the same time and the importance of such details depends on the downstream analysis to be conducted. 
We present one set of experiments on how to achieve a good balanced between enhancing contrast and preserving image details in the following sections.

\section{Experiments}
\label{sec:experiments}

\subsection{Data}
\label{sec:experiments_data}
The imaged samples were cleared liver tissue sections, as described in~\cite{morales-navarrete_liquid-crystal_2019}, stained with Phalloidin 488 antibody to label the Actin cortex at all cell borders.
Images were acquired in a Zeiss LSM 780 confocal microscope, with a Zeiss LCI Plan-Neofluar 63x 1.3 NA Gly/Water objective, a 488nm wavelength excitation laser, an emission window range of $489 - 551$nm, and a pinhole size of $1 AU$. 
Images were acquired with an isotropic voxel size of $0.3\mu m$.
The maximum imaging depth was $100 \mu m$ .

\subsection{Image Degradation Model}
\label{sec:data_degradation}
To compute synthetically degraded images, as described in Eq.~\ref{eq:degradation}, we first blur and noise each 2D slice (focal plane) using a Gaussian filter ($\sigma=20$ pixels) and Poisson noise at an estimated magnitude as described in~\cite{kalaidzidis_fluorescence_2017} (using the image analysis software \href{http://motiontracking.mpi-cbg.de }{MotionTracking}~\cite{morales-navarrete_versatile_2015}).
Synthetic images were then merged with the original images using $\alpha$ values ranging in linear steps from $0.5$ to $0.3$, with $0.5$ being used for the most superficial slice.

\subsection{Network Architecture and Hyperparameters}
\label{sec:experiments_netarchitecture}
Our network is a U-Net~\cite{ronneberger_u-net_2015}, using a depth of five, $32$ initial feature channels, an MAE loss, and a linear function as the last activation layer.
Our models were trained until convergence with an initial learning rate of $4\times10^{-4}$ for a total of $450$ epochs with $200$ steps per epoch.
A step uses a batch size of $16$, of which each patch is a $128\times 128$ pixels crop from the body of training data.
Networks were built using the \href{http://csbdeep.bioimagecomputing.com/}{CSBDeep} toolbox~\cite{weigert_content-aware_2018} using Tensorflow 2.2.1.
By default, and if not otherwise stated, we would not train the bias terms of each network node (-bias) to improve OOD predictions~\cite{mohan_robust_2020}, as also described in Section~\ref{sec:methods}.

\subsection{Baselines}
\label{sec:experiments_baselines}
Baseline methods, which we used to compare \deepcontrast with, are:
$(i)$~classical methods, \ie CLAHE and deconvolution using Huygens, and 
$(ii)$~\dl based methods, \ie the FCE-Net~\cite{zhang_fce-net_2022} trained on our liver dataset.

CLAHE images were obtained using Fiji~\cite{schindelin_fiji_2012}, where the \textit{Enhanced Local Contrast} plugin is an implementation of the original CLAHE~\cite{zuiderveld_contrast_1994} method.

Deconvolved images were obtained using the software Huygens Professional (version 22.10.0p6) from Scientific Volume Imaging (\href{http://svi.nl}{SVI}, The Netherlands), following provided pipelines with a theoretical PSF.
Internally, Huygens is using the CMLE algorithm, with SNR set to 5, for 60 iterations, and the background value set to 100.
This setup led to the best results on the data at hand.

For the FCE-Net~\cite{zhang_fce-net_2022} we used the code as it is provided by the authors of the original paper.
Its worth mentioning that the provided pre-trained FCE network led to inferior results, hence, we trained the FCE-Net from scratch until convergence using our own data.
Please note that we have also applied the FCE-Net iteratively, as described above in Equation~\ref{eq:iterations} and observed that also FCE-Net results keep improving. 
For fair comparison we are therefore reporting iterative FCE-Net results whenever they are better than single predictions.

\begin{figure*}[ht]
\begin{center}
    \includegraphics[width=1\linewidth]{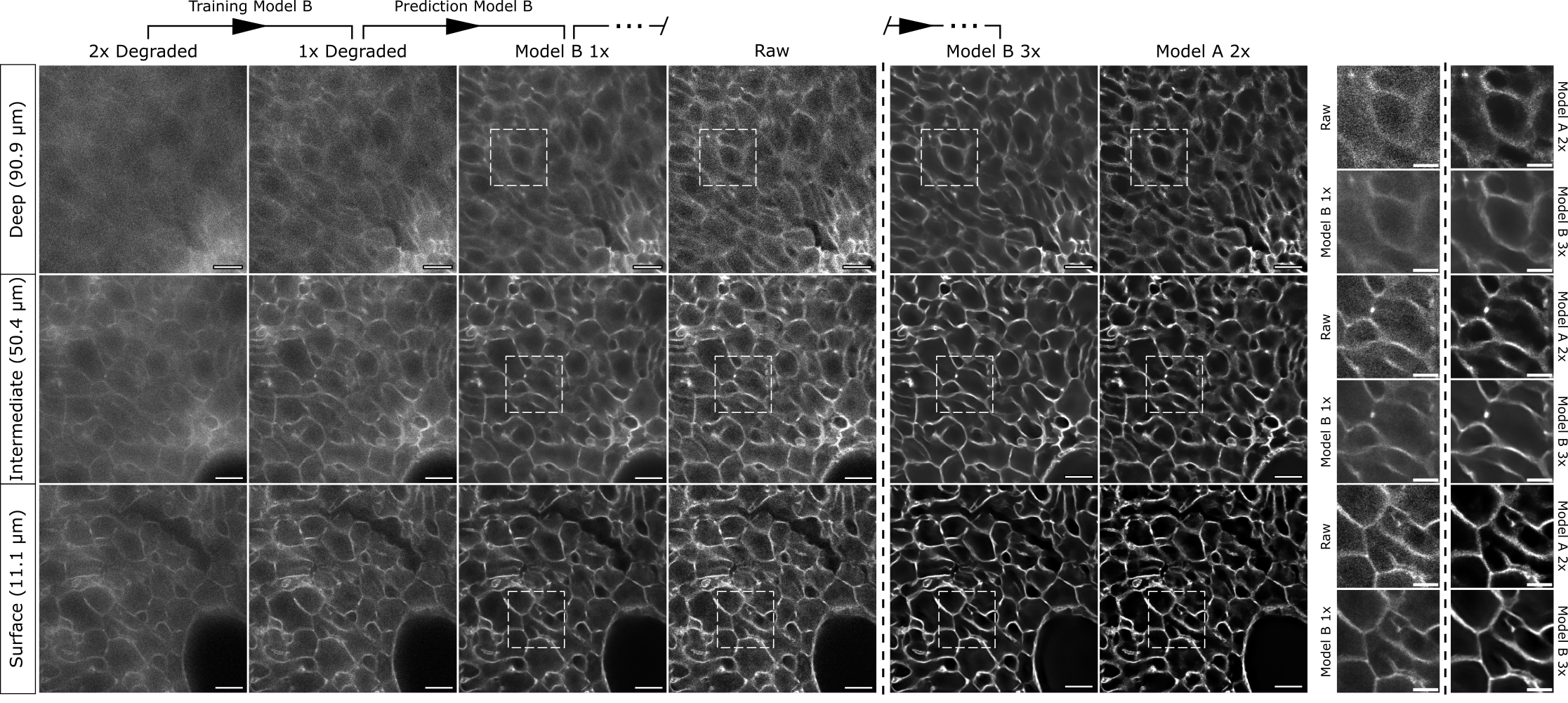}
\end{center}
    \caption{%
    Qualitative double degradation results and predictions with Model~B (see Section~\ref{sec:experiments_doubledegradation}).
    Each row shows different depths in the sample, as in previous figures.
    Columns depict, from left to right, 
    $(i)$~double degraded images used as inputs during training, 
    $(ii)$~single degraded images used as targets during training, 
    $(iii)$~predictions of Model~B when applied to data as the one in column two,
    $(iv)$~the raw image data to compare Model~B ($1\times$) outputs against,
    $(v)$~Model~B~($3\times$) outputs, and
    $(vi)$~Model~A~($2\times$) outputs to compare Model~B~($3\times$) outputs against. 
    Note that these two predictions should be and are similar since Model~B starts with inputs that are once more synthetically degraded.
    The smaller panels in the rightmost columns show the insets (marked by dashed lines) from the other columns.
    Structures in all enhanced images are consistent with structures in the raw data, which is encouraging.
    \textit{Scale bars}: 20 $\mu m$ in full size images, 10 $\mu m$ in insets.
    }
    \label{fig:double_degradation}
\end{figure*}

\begin{table}
\begin{center}
\begin{footnotesize}
    \begin{tabular}{c|c|c|c}
        \hline \hline
        & \multicolumn{3}{c}{SSIM} \\
         &\scriptsize $MB_{1\times}$ vs Raw &\scriptsize $MB_{2\times}$ vs $MA_{1\times}$ & \scriptsize $MB_{3\times}$ vs $MA_{2\times}$ \\
        \hline
        \scriptsize Very Deep & 0.51 $\pm$ 0.07 & 0.70 $\pm$ 0.08 & 0.80 $\pm$ 0.07 \\
        \scriptsize Deep & 0.52 $\pm$ 0.07 & 0.72 $\pm$ 0.09 & 0.82 $\pm$ 0.07 \\
        \scriptsize Intermediate & 0.56 $\pm$ 0.06 & 0.77 $\pm$ 0.06 & 0.84 $\pm$ 0.03 \\
        \scriptsize Shallow & 0.60 $\pm$ 0.06 & 0.80 $\pm$ 0.05 & 0.83 $\pm$ 0.03 \\
        \scriptsize Very Shallow & 0.59 $\pm$ 0.08 & 0.79 $\pm$ 0.06 & 0.83 $\pm$ 0.02 \\
        \hline
    \end{tabular}
    \vspace{1mm}
    \caption{Quantitative results of the double degradation experiment described in Section~\ref{sec:experiments_doubledegradation} and shown in Figure~\ref{fig:double_degradation}.
    We compare the outputs of three iterations of Model~B ($MB_{k\times}$), which was trained on double-degraded and single-degraded inputs, to the closest matching images, \ie raw data for direct predictions of Model~B ($MB_{1\times}$ vs Raw), direct predictions of Model~A to two iterations of Model~B ($MB_{2\times}$ vs $MA_{1\times}$), and predictions of two iterations of Model~A to three iterations of Model~B ($MB_{3\times}$ vs $MA_{2\times}$). Section~\ref{sec:methods} for details.
    }
    \label{tab:double_degrade_ssim}
\end{footnotesize}
\end{center}
\end{table}
\subsection{Double Degradation Experiments}
\label{sec:experiments_doubledegradation}
Since \deepcontrast, by definition, is applied OOD, we wondered if iterative contrast enhancement will make consistent steps.

As a first verification of our approach we degraded the raw image data twice, acquiring  data triplets $(e_i,d_i,x_i)$, with $x_i\in X$, and $d_i=\text{d}(x_i)$ and $e_i=\text{d}(\text{d}(x_i))$.
Then we trained a \deepcontrast network, Model~B, on pairs $(e_i, d_i)$ and applied the trained network, in-line with the initially proposed procedure, to $d_i$ to increase its contrast and yielding $y_i=DC(d_i)$.

Since we started by double degrading the original $x_i$, we can now compare the prediction $y_i$ with $x_i$, and further iterations of Model~B with corresponding predictions obtained with Model~A (see Section~\ref{sec:methods}).
If the trained \deepcontrast network is indeed a good approximation of the inverse of our degradation function $\text{d}$, predictions $y_i$ should be similar to the original images $x_i$ at the first iteration, and to the corresponding iteration of output images from Model~A.

\subsection{Ablations}
\label{sec:experiments_ablation}
In order to evaluate if bias-free training~\cite{mohan_robust_2020} is indeed leading to better results, we decided to repeat model training also on networks that are not bias-free, \ie train all weights and biases.

\subsection{Contrast Quantification}
\label{sec:contrast_quantification}
\subsubsection*{Wavelet Contrast Index}
\label{sec:methods_wavelet_contrast}
With increasing contrast in an image, we expect background signal to be reduced and, consequently, the brightness of biological structures in the image (signal) to be increased.
To quantify image contrast when no GT data is available, we used the Wavelet Contrast Index (WCI)~\cite{albright_paintings_2023}.
This measure computes the difference between coefficients obtained from a wavelet decomposition, following the equation
\begin{equation} \label{eq:wavelet_index}
    \text{WCI}(x) = \log(\frac{W_{95^{th}}(x)}{W_{50^{th}}(x)}),
\end{equation}
where $x$ is the input image for which we want to evaluate the contrast, $W_{95^{th}}$ is the $95^{th}$ percentile wavelet coefficient and $50^{th}$ is the median wavelet coefficient.

Wavelet decomposition was performed with the PyWavelets~\cite{lee_pywavelets_2019} python package using a Haar wavelet as the reference function and used coefficients up to the fourth level of decomposition.

\subsubsection*{Percentile Contrast Index}
\label{sec:methods_percentile_contrast}
We also used the Percentile Contrast Index (PCI) to quantify intensity differences between image background and image structures. 
The PCI is computed by
\begin{equation} \label{eq:percentile_index}
\text{PCI}(x) = \log(\frac{I_{95^{th}}(x)}{I_{50^{th}}(x)}),
\end{equation}
where $x$ is again the input image to evaluate, and $I_{95^{th}}$ is the $95^{th}$ intensity value in the image being analyzed and $I_{50^{th}}$ is the median intensity of $x$.
We use the median value, assuming that at least half the pixels of any given image are background pixels.

\subsection{Downstream Segmentation after Contrast Enhancement}
\label{sec:downstream_analysis}
Enhancing contrast not necessarily improves downstream process-ability (interpretability) of a given dataset.
While the contrast, as measured by WCI and/or PCI, might still improve, details relevant for biological interpretation of the data might already get lost.
Therefore, the best amount of contrast enhancement depends on a given downstream analysis task.
To this end, we introduced a downstream segmentation task and checked if a fixed segmentation pipeline improved with respect to existing ground truth labels.
GT segmentation masks were generated from raw data using Labkit~\cite{arzt_labkit_2022} (available as a Fiji~\cite{schindelin_fiji_2012} plugin).

For simplicity, we segmented contrast enhanced images $y_i$ by thresholding, optimizing for the best threshold value, \ie the one that maximizes the intersection-over-union (IoU) with respect to the previously generated GT. 
Our reasoning was that enhancing contrast would result in a better IoU after thresholding as long as relevant structures in the contrast enhanced images $y_i$ were not lost.
As soon as details were getting lost, the IoU dropped, allowing us to choose the most sensible iteration depth for \deepcontrast (or the FCE-Net).

\section{Results}
Qualitative results presented in Figure~\ref{fig:results_qualitative} suggest that \deepcontrast outperforms all baseline methods.
\deepcontrast removes or reduces image noise and enhances the intensity of visible image structures seemingly without loosing fine details (signal) from predicted images. 
Hence, \deepcontrast is indeed increasing image contrast.

Both classical methods, \ie CLAHE and deconvolution, displayed relatively poor results mainly deep into the tissue.
CLAHE amplified image noise at all imaging depths and mostly failed to highlight biological structures.
Deconvolution, on the other hand, did reduce image noise, but failed to increase intensities of foreground structures (most obvious deep into the tissue). 

The FCE-Net performed much better, leading to good results close to the surface. 
But the image contrast in FCE-Net predictions decayed with increasing depth (see insets in Figure~\ref{fig:results_qualitative}). 
Qualitatively, the FCE-Net also seemed to produce less sharp cell borders (as seen in either deep and shallow image regions).

To validate these qualitative observations, we quantified contrast with the two measures WCI and PCI (see Section~\ref{sec:contrast_quantification}).
As can be seen in Figure~\ref{fig:results_quantitative}, \deepcontrast achieved higher image contrast at all imaging depths and over all plotted iterative applications ($1\times$ to $3\times$).
One notable exception are the WCI values for $3\times$ iterations in intermediate imaging depths.
In these images, despite the FCE-Net showing higher WCI values, one can see more image structure being lost in FCE-Net predictions than in predictions obtained with \deepcontrast (see Figure~\ref{fig:multi_segmentation} for a qualitative and quantitative comparison).

As introduced above, contrast enhancement can be applied iteratively (Equation~\ref{eq:iterations}).
Results of performing multiple rounds of enhancement are shown in Figure~\ref{fig:iterative_inference}.
Visually, the best results were obtained with three rounds of enhancement ($3\times$).
While contrast readouts using WCI and PCI would still improve with additional iterations, image details would start disappearing (as can be seen in the $6\times$ column and the line-plots in Figure~\ref{fig:iterative_inference}).

Qualitative results of the Double Degradation Experiments introduced in Section~\ref{sec:experiments_doubledegradation}, are shown in Figure~\ref{fig:double_degradation}.
Predictions of Model~B at iteration $k$ should and are corresponding well to predictions of Model~A at iteration $k-1$ since Model~B is trained on image pairs that are one application of our forward degradation model ($\text{d}$) more degraded.
We quantify this via structure similarity index measure (SSIM)~\cite{wang_image_2004} in Table~\ref{tab:double_degrade_ssim} and allow for a qualitative comparison between the corresponding columns in Figure~\ref{fig:double_degradation}.

\subsection{Contrast Enhancement vs.\ Segmentation}
To better quantify the undesired effect of loosing relevant details while simultaneously gaining additional contrast in the processed microscopy data, we introduced a simple threshold based segmentation task (See Section~\ref{sec:downstream_analysis}).
A qualitative as well as quantitative comparison is shown in Figure~\ref{fig:multi_segmentation}.
IoU values are initially increasing with number of iterations, but then eventually drop when too many image structures are removed. 
The FCE-Net generally shows lower IoU values, suggesting that \deepcontrast is not only leading to more contrasted images, but is at the same time maintaining more image details with iterations. 
In addition to the IoU quantification, we highlighted lost details on images with yellow arrow heads  (see in Figure~\ref{fig:multi_segmentation} ), pointing differences between iterative inferences.

\subsection{Ablation: Training including Bias}
\begin{figure}[ht]
\begin{center}
    \includegraphics[width=1\linewidth]{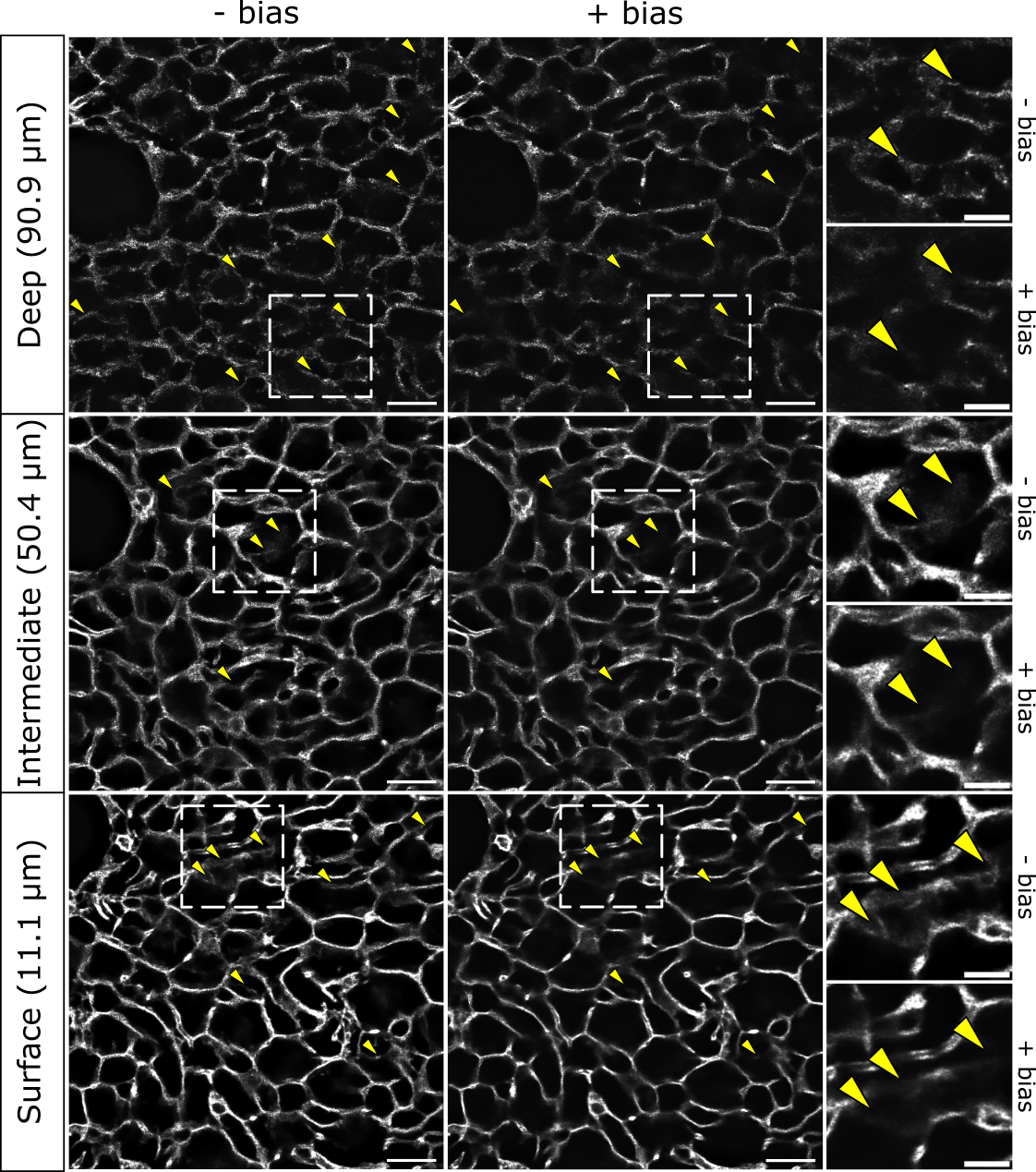}
\end{center}
    \caption{%
    Qualitative results of networks trained without and with bias.
    Phalloidin stained images of liver tissue sections enhanced $3\times$ with \deepcontrast models trained with (+bias; right side) and without bias (-bias; left side).
    Rows show image planes at different depths relative to cover-glass.
    Network model trained with bias performs worse when applied OOD, removing structures seen in a model trained without bias (- bias), highlighted by yellow arrow-heads.
    \textit{Scale bars}: 20 $\mu m$ in full size images, 10 $\mu m$ in insets.
    }
    \label{fig:results_bias_vs_nobias}
\end{figure}

As introduced above, \deepcontrast employs bias-free~\cite{mohan_robust_2020} network training.
In Figure~\ref{fig:results_bias_vs_nobias} we show representative predictions of Model~A ($3\times$), as used in Figure~\ref{fig:results_qualitative}, and compare them to predictions obtained with an equivalent model which was trained with bias (+bias). 
Yellow arrow heads in the figure point at locations where the bias free network does a better job retaining image details.
Empirically, we did not spot any cases where the opposite is true, which gave us additional motivation to use bias-free networks in \deepcontrast.

\section{Discussion and Conclusion}
\label{sec:discussion}
In this work we propose to use an image degradation function to approximate light scattering in deep tissue imaging and use it to generate synthetically degraded data to enable supervised network training.
Our results show that the relatively simple degradation model we introduced is sufficient to increase image contrast in real microscopy data.
Our method can be applied in an iterative manner to further increase image contrast and will retain detailed image structures for more iterations than the competitive baseline methods we compared against.

For the liver data at hand, we have found that the best number of iterations for contrast enhancement is three ($3\times$).
This assessment is based on a combination of contrast enhancement and retention of fine image details in the contrast enhanced predictions.
A more quantitative approach to the visual assessment was introduced by means of a downstream segmentation task, which has indeed confirmed our initial findings.

In general, the best trade-off between contrast enhancement and structural integrity of predictions depends on the nature of the downstream processing task to be conducted.
Hence, an analysis similar to the one we performed for the segmentation task could be required to evaluate the best-performing setup.

Similarly we found that for better OOD application of our trained networks, the bias free version seems to lead to better results.

While our approach is leading to excellent results and can easily be used by microscopists and life scientists to improve volumetric image data for quantitative downstream processing, additional research will be required to undo image degradations deep in imaged tissues in more fundamental ways.

\section*{Acknowledgments}
The authors thank José Valenzuela-Iturra for acquiring image data.
We thank the LMF and SCF facilities at MPI-CBG for technical support.
We thank Igor Zubarev for help and feedback with the presented experimental setups.
We thank Ashesh, Anirban Ray, Sheida Rahnamai Kordasiabi, Igor Zubarev, Joran Deschamps and Damian Dalle Nogare for helpful discussion and feedback.
This work was supported by the European Research Council ERC Advanced Rulliver Grant (no. 695646) to M. Zerial.
Additionally, this work was supported by 
the European Union through the Horizon Europe program AI4LIFE with grant agreement 101057970-AI4LIFE.
Funding was also provided from the
Max-Planck Society under project code M.IF.A.MOZG8106.

{\small
\bibliographystyle{ieee_fullname}
\bibliography{references}
}

\end{document}